\documentclass[12pt]{iopart}

\usepackage{amsfonts,amssymb}
\usepackage{graphicx}
\usepackage[american]{babel}

\begin{document}

\title{A multiscale view on inverse statistics and gain/loss asymmetry in financial time series}
\author{Johannes Siven$^1$, Jeffrey Lins$^1$ and Jonas Lundbek Hansen$^1$}
\address{$^1$ Saxo Bank A/S, Philip Heymans All\'e 15, DK-2900 Hellerup, Denmark}
\ead{\mailto{jvs@saxobank.com}, \mailto{jtl@saxobank.com}}
\begin{abstract}
Researchers have studied the \emph{first passage time} of
financial time series and observed that the smallest time interval
needed for a stock index to move a given distance is typically
shorter for negative than for positive price movements. The same
is not observed for the index constituents, the individual stocks.
We use the discrete wavelet transform to illustrate that this is a
long rather than short time scale phenomenon
--- if enough low frequency content of the price process is
removed, the asymmetry disappears. We also propose a new model,
which explain the asymmetry by prolonged, correlated down
movements of individual stocks.
\end{abstract}

\section{Introduction}
\begin{figure}
\begin{center}
\includegraphics[width=8cm]{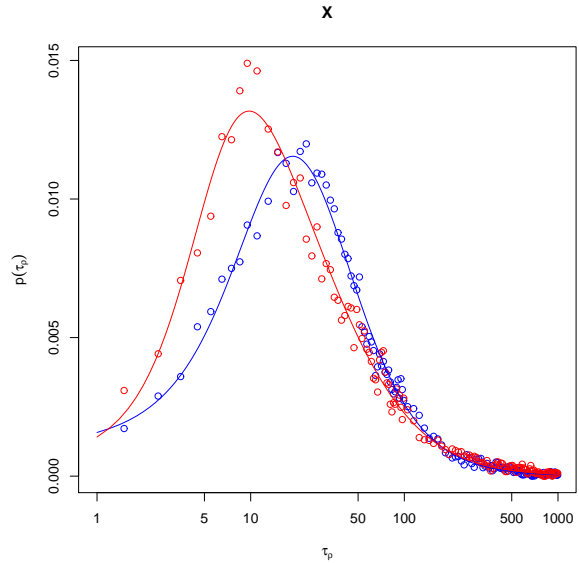}
\end{center}
\caption{Estimated distribution of the first passage time
$\tau_\rho$ for the log price of the Dow Jones industrial Average
index. The graphs correspond to $\rho = +5\%$ (blue) and $\rho =
-5\%$ (red).} \label{fig:DJIA_FPT}
\end{figure}

Modelling the statistical properties of financial time series has
long been an active area of research, both in the spaces of
economics and physics. The traditional object of study has been
the returns of various assets, i.e.\ the size of price movements
over fixed time intervals. Inspired by research in the field of
turbulence, Simonsen, Jensen and Johansen \cite{investHorizon}
asked the ``inverse" question: what is the smallest time interval
needed for an asset to cross a fixed return level $\rho$? Figure
\ref{fig:DJIA_FPT} shows the distribution of this random variable,
the \emph{first passage time}, for the Dow Jones Industrial
Average index, for $\rho = \pm 5\%$. As noted by Jensen, Johansen
and Simonsen \cite{investStatistics}, the most likely first
passage time is shorter for $\rho = -5\%$ than for $\rho = 5\%$,
which they refer to as the \emph{gain/loss asymmetry}.
Intriguingly, the same asymmetry is \emph{not} observed in the
constituents of the index, the individual stocks
\cite{inverseMarkets}.

One explanation for the gain/loss asymmetry in the index is that
occasionally, the stocks move in a highly correlated manner, and
that this tend to happen for down movements rather than up
movements \cite{fearModel}. To support this view, Donangelo,
Jensen, Simonsen and Sneppen \cite{fearModel} proposed the
\emph{asymmetric synchronous market model}, where the log stock
prices move like independent random walks for most days, but
sometimes move downwards together. This model exhibits gain/loss
asymmetry for the index, but not for the individual stocks.

In this paper we investigate \emph{on what time scale} the
gain/loss asymmetry emerges. To this end, we use the discrete
wavelet transform and estimate the first passage time for
high-pass filtrations of the time series. Apparently, if enough
low frequency content is removed, the gain/loss asymmetry
disappears
--- the asymmetry is due to effects on quite long time scales,
64-128 days and longer. Together with the fact that individual
stocks do not exhibit gain/loss asymmetry this indicates that the
asymmetry is due to prolonged, correlated down-movements of the
stock prices. This is contrary to the model from Donangelo et al.\
\cite{fearModel}, where the correlated losses are ``local" in
time. Indeed, we illustrate that the gain/loss asymmetry in that
model emerges on a rather short time scale, which is inconsistent
with our empirical findings. Finally, we construct a new model
where the gain/loss asymmetry emerges on longer time scales,
consistent with the empirical findings.

\section{First passage time on multiple time scales}
Throughout this section we let $\{X_t\}_{t \geq 0}$ denote the
logarithm of a financial time series, for instance the price of a
stock index. The first passage time $\tau_\rho$ of the level
$\rho$ is defined as
$$
\tau_\rho = \left\{ \begin{array}{ll}
\inf\{s>0;\ X_{t+s} - X_t \geq \rho\}& \mbox{if } \rho> 0,\\
 \inf\{s>0;\ X_{t+s} - X_t \leq \rho\}& \mbox{if } \rho< 0,
\end{array} \right.
$$
and is assumed to be independent of $t$. The distribution of
$\tau_\rho$ is estimated in a straight forward manner from a time
series $X = (X_{t_1},\ldots,X_{t_N})$. Consider $\rho > 0$, and
let $t_{n+m}$ be the smallest time point such that $X_{t_{n+k}} -
X_{t_n} \geq \rho$, if such a time point exists. In that case,
$t_{n+m} - t_n$ is viewed as an observation of $\tau_\rho$. (If
$\rho <0$, take instead $t_{n+m}$ such that $X_{t_{n+k}} - X_{t_n}
\leq \rho$.) Running $n$ from $1$ to $N-1$ gives a set of
observations from which the distribution of $\tau_\rho$ is
estimated as the empirical distribution.

Given a time series $X = (X_{t_1},\ldots,X_{t_N})$, the level-$J$
discrete wavelet transform\footnote{See, e.g., Gencay, Selcuk and
Whitcher \cite{Whitcher} or Percival and Walden \cite{Percival}.}
yields an additive decomposition
$$
X = \sum_{j = 1}^{J}D_j + S_J,
$$
where $D_j$ is the $j$:th level \emph{detail} and $S_J$ is the
$J$:th level \emph{smooth}. Essentially, $D_j$ is a band-pass
filtration and $S_J$ is a low-pass filtration of $X$. If $X$
consists of daily observations, then $D_j$ contains changes on a
time scale of between $2^{j-1}$ and $2^{j}$ days, and $S_J$
contains changes on timescales longer than $2^{J}$ days. The
signal $R_J := X - S_J$ can thus be seen as a ``detrended" version
of $X$, where the time horizon of the removed trends increases
exponentially with $J$.

\begin{figure}
\begin{center}
\includegraphics[width=8cm]{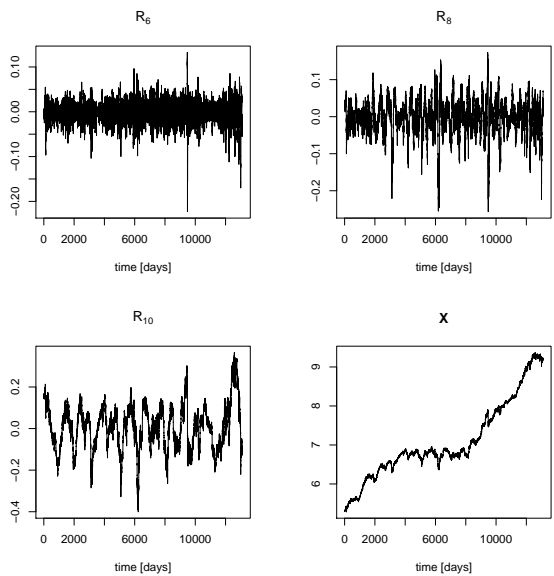}
\end{center}
\caption{The log price of the Dow Jones Industrial Average index,
$X$, and its high-pass filtrations $R_6,R_8$ and $R_{10}$.}
\label{fig:DJIA_MRA}
\end{figure}

\begin{figure}
\begin{center}
\includegraphics[width=8cm]{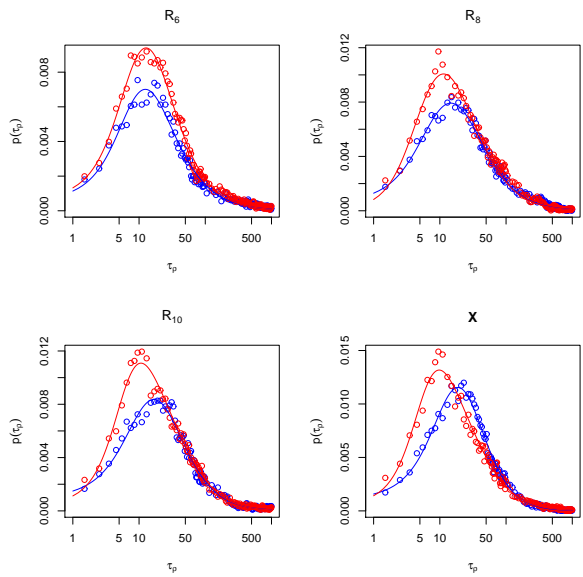}
\end{center}
\caption{Estimated distribution of the first passage time
$\tau_\rho$ for $X$, the log price of the Dow Jones Industrial
Average index, and its high-pass filtrations $R_6, R_8$ and
$R_{10}$. The graphs correspond to $\rho = +5\%$ (blue) and $\rho
= -5\%$ (red).} \label{fig:DJIA_MRA_FPT}
\end{figure}

We considered the time series $X$ of daily observations of the Dow
Jones Industrial Average index, henceforth DJIA, and
computed\footnote{In all our experiments we use LA(8), the
\emph{least asymmetric} wavelet filter of length 8, see Daubechies
\cite{LA8}, but our results are robust in the sense that other
choices yield very similar results.} the filtered signals
$R_6,R_8$ and $R_{10}$ (see Figure \ref{fig:DJIA_MRA}). We then
estimated the distribution of the first passage time for
$R_6,R_8,R_{10}$, and $X$. Figure \ref{fig:DJIA_MRA_FPT} show the
result: the gain/loss-asymmetry is absent for $R_6$, and gradually
emerges for $R_8$ and $R_{10}$. This indicates that if enough low
frequency components of the signal are removed, the gain/loss
asymmetry disappears. Since the level $J = 6$ corresponds to 32-64
days, the asymmetry appears mainly due to effects on rather long
time scales: 64-128 days and longer. The same pattern was observed
for other stock indices, like the S\&P500 and Nasdaq (not
reported).

\begin{figure}
\begin{center}
\includegraphics[width=8cm]{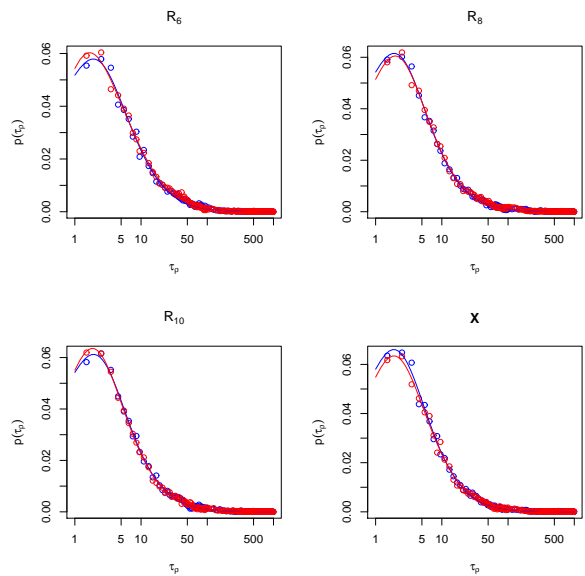}
\end{center}
\caption{Estimated distribution of the first passage time
$\tau_\rho$ for $X$, the log price of the IBM stock, and its
high-pass filtrations $R_6, R_8$ and $R_{10}$. The graphs
correspond to $\rho = +5\%$ (blue) and $\rho = -5\%$ (red).}
\label{fig:IBM_MRA_FPT}
\end{figure}

Figure \ref{fig:IBM_MRA_FPT} shows the results from the same
analysis for the IBM stock price. We see that there is virtually
no asymmetry for the IBM stock, at any time scale, which is
consistent with the observations from Johansen et al.\
\cite{inverseMarkets}. We have also confirmed this for several
other individual stocks.

\begin{figure}
\begin{center}
\includegraphics[width=8cm]{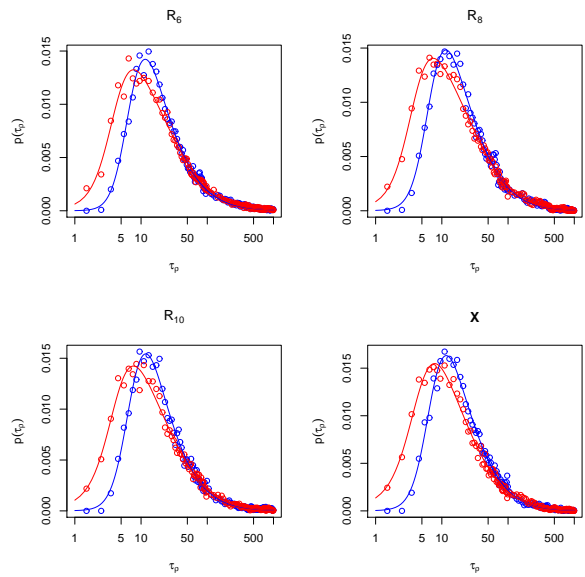}
\end{center}
\caption{Estimated distribution of the first passage time
$\tau_\rho$ for $X$, the log price in a realization of ASMM with
$N = 30$ stocks, and its high-pass filtrations $R_6,R_8$ and
$R_{10}$. The graphs correspond to $\rho = 5 \sigma$ (blue) and
$\rho = -5 \sigma$ (red).} \label{fig:ASMM_MRA_FPT}
\end{figure}

In light of the finding that the gain/loss asymmetry in stock
indices are due to effects on long rather that short time scales,
it is natural to question the structure of the asymmetric
synchronous market model (ASMM) given by Donangelo et al.\
\cite{fearModel}. In that model, the features giving rise to the
gain/loss asymmetry by construction takes place on the shortest
possible time scale. To see to what extent the gain/loss asymmetry
in ASMM resembles that of DJIA when considered on multiple time
scales, we performed the analysis for a realization of the model.
That is, we let $X$ be a time series generated by ASMM, and
estimated the first passage time densities\footnote{As in
Donangelo et al.\ \cite{fearModel}, we take $\rho = \pm 5\sigma$
where $\sigma$ equals the daily volatility (the standard deviation
of the log price) of the model index. This corresponds to $\rho =
\pm 5\%$ in the case of the Dow Jones index, since the daily
volatility there is roughly 1\%.} for $R_6, R_8, R_{10}$, and $X$.
Figure \ref{fig:ASMM_MRA_FPT} shows the result
--- the gain/loss symmetry is evidently present already on shorter
time scales, for
 $R_6$ and $R_8$, which is in contrast with the empirical findings (cf.\ Figure
\ref{fig:DJIA_MRA_FPT}).

\section{A new model}
In this section we propose a new model to remedy the failure of
ASMM in reproducing the gain/loss asymmetry when considered on
multiple time scales. The new model can be seen as a
generalization of ASMM --- it has one ``regular" state where the
stocks move independently, and one ``distressed" state where their
moves are highly correlated. The most important difference between
our model and ASMM is that the market may remain in the distressed
state for several days. Conceptually, this is in correspondence
with the empirical findings from the previous section, that the
gain/loss asymmetry is due to effects on long rather than short
time scales.

In the regular state, all stocks follow independent geometric
Brownian motions with drifts $\mu_r$ and standard deviation $\xi$:
$$
S_i(t+\Delta t) = S_i(t) e^{(\mu_r - \frac{\xi^2}{2})\Delta t +
\xi \sqrt{\Delta t} Z_i},
$$
where $Z_i$, $i = 1,\ldots,N$, are independent standard normals.
In the distressed state, there is a pronounced negative drift
$\mu_d < 0$, and all the stocks move together:
$$
S_i(t+1) = S_i(t) e^{(\mu_d - \frac{\xi^2}{2})\Delta t + \xi
\sqrt{\Delta t} Z},
$$
where $Z$ is standard normal --- the single random variable $Z$
drives the price change in all the stocks. For any given day, we
consider the possibility of changing states: we let $p_{rd}$
denote the probability of changing from the regular to the
distressed state, and let $p_{dr}$ denote the probability of the
converse transition. The index is defined by
$$
I(t) := \frac{1}{N}\sum_{i = 1}^N S_i(t).
$$
Note that ASMM can be seen as a particular case of this model,
with $p_{dr} = 1$ --- the transition from the distressed to the
normal state always happens in one day.

We fix the following parameters: $\Delta t = 1/250$, $\xi = 0.3$,
$\mu_d = -0.15$, $p_{rd} = 1/200$, $p_{dr} = 1/25$, and choose the
drift $\mu_r$ in order to make $\mathbb{E}[(I(t+\Delta t) -
I(t))/I(t)] = 2.58\cdot 10^{-4}$ (this is the historical daily
mean return of DJIA). As above, let $\rho = \pm 5 \sigma$, where
$\sigma$ denotes the daily standard deviation of the change in
$\log(I(t+\Delta t)/I(t))$.

\begin{figure}
\begin{center}
\includegraphics[width=8cm]{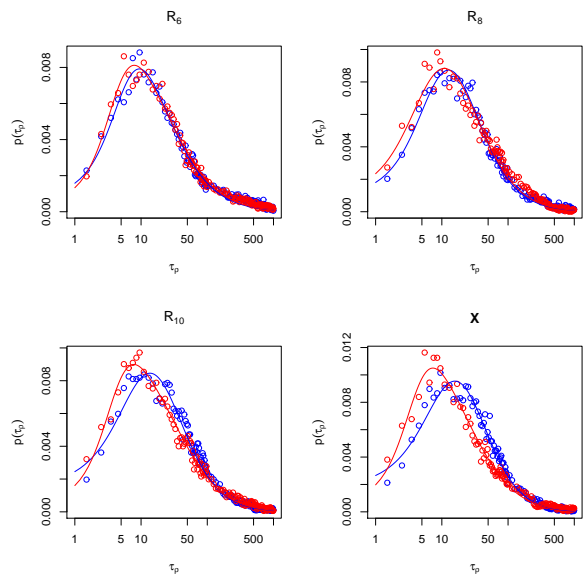}
\end{center}
\caption{Estimated distribution of the first passage time
$\tau_\rho$ for $X$, the log price in a realization of our
proposed model with $N = 30$ stocks, and its high-pass filtrations
$R_6,R_8$ and $R_{10}$. The graphs correspond to $\rho = 5 \sigma$
(blue) and $\rho = - 5\sigma$ (red).} \label{fig:MRA_NewModel30}
\end{figure}

\begin{figure}
\begin{center}
\includegraphics[width=8cm]{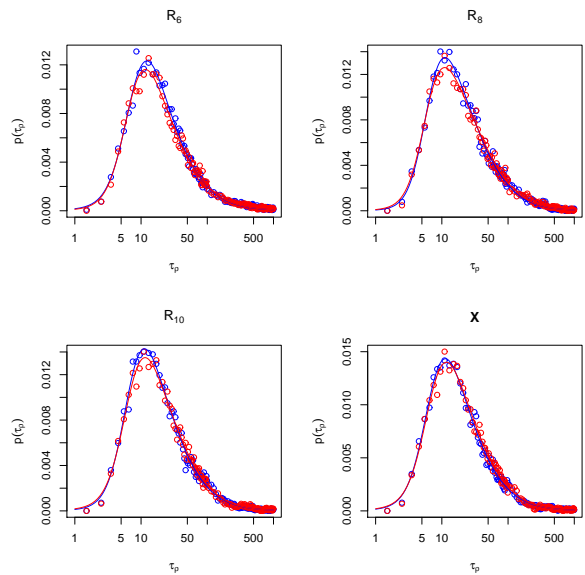}
\end{center}
\caption{Estimated distribution of the first passage time
$\tau_\rho$ for $X$, the log price in a realization of our
proposed model with $N = 1$ stock, and its high-pass filtrations
$R_6,R_8$ and $R_{10}$. The graphs correspond to $\rho = 5 \sigma$
(blue) and $\rho = - 5\sigma$ (red).} \label{fig:MRA_NewModel1}
\end{figure}

Figure \ref{fig:MRA_NewModel30} show the estimated first passage
density for the index and its high-pass filtrations, for $N = 30$
stocks. We see that the multiresolution analysis resembles that of
DJIA: the gain/loss asymmetry is absent for $R_6$ but emerges for
$R_8$ and $R_{10}$. Figure \ref{fig:MRA_NewModel1} show the result
from the same analysis but with $N = 1$ stock: as expected, there
is no gain/loss asymmetry in this case. Again, this is consistent
with the empirical findings, that individual stocks do not exhibit
gain/loss asymmetry.

\newpage
\section{Conclusion}
Using wavelet multiresolution analysis, we have shown that the so
called gain/loss asymmetry observed in stock indices is due to
effects on long rather than short time scales --- the asymmetry
disappears if enough low frequency content is removed from the
signal. Moreover, inspired by the asymmetric synchronous market
model, we proposed a new model featuring prolonged periods of
correlated down movements of the index constituents that
qualitatively reproduces the asymmetry.

According to our new model, the stock market occasionally enters a
distressed state of correlated down movements and stays there for
a prolonged period of time, expected to last $1/p_{dr}$ days. In
the future, we would like to investigate whether it is possible to
estimate this quantity from data by comparing empirical first
passage time distributions to the distributions implied by the
model, in effect ``calibrating'' the model. We are intrigued by
the question of how such information may relate to the so called
\emph{Heterogeneous Market Hypothesis} \cite{HFfinance}
--- that is, that various participants in the market have different
time horizons and dealing frequencies. The wavelet multiscale
decomposition of the signal could potentially be used be to
investigate the emergence of the gain/loss asymmetry even further.
For instance, what happens if the signal is band-pass instead of
high-pass filtered --- is it possible to pinpoint the time scale
of the origin of the asymmetry?

\section*{References}
\bibliographystyle{plain}
\bibliography{JSbib_inverse}

\end{document}